\documentclass[12pt]{article}
\usepackage[dvips]{color}
\usepackage{amsmath}
\usepackage{graphicx}
\usepackage{subfigure}
\usepackage{epsfig}
\usepackage{amsmath}
\usepackage{cite}
\usepackage{color}
\usepackage{subfigure}

\begin{document}
\begin{center}
\Large{\bf Can black holes cause cosmic expansion?}\\
\small \vspace{1cm}
{\bf J. Sadeghi$^{\star}$\footnote {Email:~~~pouriya@ipm.ir}}, \quad
{\bf S. Noori Gashti $^{\star,\dag}$\footnote {Email:~~~saeed.noorigashti@stu.umz.ac.ir}}, \quad
{\bf M. R. Alipour$^{\star}$\footnote {Email:~~~mr.alipour@stu.umz.ac.ir}}
and\quad{\bf M. A. S. Afshar $^{\star}$\footnote {Email:~~~m.a.s.afshar@gmail.com}}\\
\vspace{0.5cm}$^{\star}${Department of Physics, Faculty of Basic
Sciences,\\
University of Mazandaran
P. O. Box 47416-95447, Babolsar, Iran}\\
\vspace{0.5cm}$^{\dag}${School of Physics, Damghan University, P. O. Box 3671641167, Damghan, Iran}
\small \vspace{1cm}
\end{center}
\begin{abstract}
Recently, it was shown in \cite{1} that black holes are the source of dark energy. In \cite{2,3,4,cc}, the truth or falsity of this concept has been discussed. We briefly state the arguments raised in each of these papers, but our main goal is not to accept or reject these debates. Rather, in this note, we show that black holes in specific structures, such as the Reissner-Nordstrom (R-N) black hole in the presence of quintessence, string cloud, and perfect fluid, can produce a repulsive force with respect to the weak gravity conjecture mechanism that overcomes the gravitational force.\textit{ We discuss these repulsions forces in a population of black holes of the same family can be considered as an auxiliary factor (albeit a weak factor) for the expansion of the universe}.
\end{abstract}
\newpage
\section{Warm-up the Mind: WGC, Black holes, RFC $\&$ Dark energy}
The connection between black holes and the modern hypothesis of the weak gravity conjecture (WGC) and its motivations can be examined from several perspectives. Meanwhile, perhaps the first and most important form of this relationship can be expressed as follows: any subextremal black hole should be able to decay into lighter charged particles and avoid becoming overextremal, which would violate cosmic censorship.
Or, from another point of view, it can be said that the weak gravity conjecture suggests the existence of superextremal black holes with charge-to-mass ratios greater than one. These black holes have a naked singularity, and their thermodynamics and stability are not well understood. In other words, the WGC implies that any particle carrying a gauge charge should have a mass smaller than its charge in Planck units. This means that such particles can always escape from a black hole by emitting gauge bosons.
There are extremal black holes with charge-to-mass ratios larger than one, which are called superextremal black holes. The Weak Gravity Conjecture (WGC) implies that superextremal black holes should not exist in nature. However, some studies have suggested that superextremal black holes may exist in anti-de Sitter (AdS) space, which is a negatively curved spacetime that arises in some models of quantum gravity. These studies have argued that superextremal AdS black holes may be stable against quantum decay and may not violate the WGC if there are additional effects such as angular momentum or higher-dimensional corrections. Therefore, the concepts are true if the WGC is valid and if there are no such effects that allow superextremal AdS black holes to exist.
And finally, it can be mentioned that the WGC shows that there must be extreme black holes with mass ratios equal to one, which are called BPS black holes. These black holes maintain some degree of supersymmetry and have a special relationship between their entropy and their charges. They also satisfy some geometric conjectures related to Calabi-Yau manifolds \cite{53,54,55,56,57}.\\\\
The WGC can be extended to include non-Abelian gauge fields and scalar fields, which can affect the properties of black holes in different ways. For example, scalar fields can change the charge-to-mass ratio of outer black holes, and non-Abelian gauge fields can induce phase transitions between different types of black holes.
Of course, it must be acknowledged that a practical investigation of the WGC, in reality, would be a very challenging task, as it would mean probing very high energy scales and very small distances where quantum gravity effects are expected to be important.

But it should be noted that the above statement does not necessarily mean that the WGC is only a hypothetical, theoretical, and impractical theory. There are some possible ways to test the WGC indirectly or in some special cases. One possibility is to look for extremal black holes with WGC conditions on their charge-to-mass ratio. Such black holes would be too unstable to become particles that satisfy the WGC and thus should not exist in a consistent theory of quantum gravity. However, finding such black holes requires very precise measurements of their mass and charge, as well as a good understanding of the environment and possible sources of error.
In fact, the basis of the WGC in the theory of quantum gravity is the comparison of the strength of gravity with the gauging forces in the studied structure, which implicitly states that gravity should be the weakest force. Of course, it should be stated that the relationship between the repulsive force and the weak gravity conjecture is not yet fully understood. Still, among the studies, we can point to significant theories that state that there are at least two inequivalent forms of the weak gravity conjecture. One is based on the charge-to-mass ratio, and the other is based on long-range forces. The electric and magnetic WGC are two forms of the WGC that apply to a U(1) gauge theory coupled to gravity \cite{53,54,55,56,57}.\\\\
The electric WGC states that there is a charged particle whose charge-to-mass ratio is at least as large as that of an extremal black hole. This means that such a particle can always escape from a black hole by emitting gauge bosons. The electric WGC is motivated by the requirement for the absence of stable black holes, which would violate the second law of thermodynamics and the unitarity of quantum gravity.
The magnetic WGC states that there is a charged particle that repels itself when far apart. This means that such a particle has a self-interaction energy that is larger than its gravitational energy. The magnetic WGC is motivated by the requirement for the consistency of quantum field theory in curved spacetime, which implies that the effective coupling of the gauge field should run to zero at large distances.
The electric and magnetic WGC are not equivalent, and they may have different implications for various aspects of quantum gravity and cosmology. For example, the electric WGC constrains the spectrum of axions and instantons, which are scalar fields and (0+1)-dimensional branes coupled to a 2-form gauge field. The magnetic WGC constrains the spectrum of strings and domain walls, which are (1+1)-dimensional and (2+1)-dimensional branes coupled to a 2-form gauge field. These constraints may affect the viability of large-field inflation models based on axions or strings.\\\\
The WGC could have implications for dark energy, which is a mysterious form of energy of unknown nature and origin that exerts negative pressure and causes the faster expansion of space-time. This is because it constrains the possible values of the cosmological constant and the equation of state of dark energy.
The cosmological constant is a term in Einstein's equations that represents a constant energy density filling space and causing its accelerated expansion. The WGC implies that the cosmological constant cannot be too large or too negative; otherwise, it would violate the bound on the charge-to-mass ratio of particles. This could help to explain why the observed value of the cosmological constant is very small but positive.
Another way in which the WGC could help is by constraining the possible equation of state of dark energy, which is a parameter that characterizes how the pressure of dark energy relates to its energy density. The WGC implies that the equation of state of dark energy cannot be too negative; otherwise, it would lead to instabilities or inconsistencies in quantum gravity theories. This could help to rule out some exotic models of dark energy that predict a very negative equation of state \cite{53,54,55,56,57}.\\\\
A third approach to constraining the possible interactions between dark energy and other forms of matter, such as dark matter, is through the WGC. The WGC is a hypothetical principle that posits that gravity should be the weakest force in nature. It implies that any particle carrying a gauge charge should have a mass smaller than its charge in Planck units, allowing such particles to escape from a black hole by emitting gauge bosons.
If dark energy is modeled by a scalar field that couples to dark matter, the WGC suggests that the coupling strength should be less than the gravitational strength. Otherwise, dark matter particles would have a mass larger than their charge in Planck units, violating the WGC bound. This would also indicate that the interaction between dark energy and dark matter is stronger than gravity, leading to various cosmological issues. Thus, if the WGC is valid and dark energy and dark matter are coupled by a scalar field, this statement holds.
However, these are not proven facts but hypotheses based on theoretical considerations and observational limitations. This approach could assist in testing some dark energy models that incorporate plausible phenomenological interactions between dark particles.\\\\
The above concepts are related by the idea that quantum gravity imposes constraints on the strength and nature of forces in a consistent theory. The weak gravity conjecture (WGC) and the repulsive force conjecture (RFC) express these constraints based on different aspects of long-range interactions between particles. Both conjectures have implications for the consistency of various models of quantum gravity, such as string theory, as well as for cosmology and particle physics. However, both conjectures face challenges, such as finding a precise formulation that applies to general situations, such as black holes with multiple charges or higher-dimensional spacetimes. There are also different versions of the WGC and the RFC, such as the strong WGC or the finite-size RFC, that impose different or stronger constraints on the spectrum and interactions of particles in a quantum gravity theory. Finally, although some of these versions have been proven or disproven in specific cases, a general proof or disproof of the WGC and RFC remains elusive\cite{53,54,55,56,57}.

\section{Results review of \cite{1,2,3,4,cc}}
Inflation and cosmic expansion are deep and not-so-old concepts whose existence we can understand and even evaluate in the universe. Despite all calculations and efforts, the origin of their existence has been considered a mysterious issue, and every attempt to decipher it has been unsuccessful until today. Recently, new approaches have emerged towards decoding this missing link, like pieces of a puzzle. In these few years, physicists have witnessed the birth and formation of an amazing triangle to decipher the nature and expansion of the universe. At the three vertices of this triangle, there emerge concepts that are not yet well-known completely to us, but finding the connection and combining their concepts may be the missing key! (Black Holes, Dark Matter, and Dark Energy).
Efforts to justify the existence of dark matter and dark energy using fundamental black hole concepts are emerging. Of course, the connection between black holes and dark matter is not far from the mind because both have mass, and evidence such as the absence of baryons in black holes can prove this. However, it seems a little more difficult to find the connection between black holes and dark energy.\\\\
In early 2023, it was suggested that the origin of dark energy is associated with black holes that have a constant effective energy density \cite{1}. The authors of the paper implied that the mass of black holes increases over time according to the same law as the volume of the part of the universe that they occupy, and therefore the population of black holes behaves similarly to dark energy. However, this assumption is not very stable and has been questioned \cite{2,3}. Additionally, it is not necessarily possible to ensure the expansion of the universe based solely on the density of black holes. Experimental research in \cite{1} suggests that the remaining stellar black holes are the astrophysical source of dark energy, explaining the accelerated expansion. It should be noted that other alternatives have been proposed, such as gravitational stars (Gravitational vacuum condensate stars = gravastars), which have been considered as an alternative to black holes and as a possible astrophysical source of dark energy. However, this option does not meet the necessary and sufficient conditions to achieve this position \cite{3}.
The authors in \cite{3} make an important point: "gravastars cannot be a source of dark energy, and we emphasize that direct coupling of their mass to the dynamics of the universe can lead to an additional acceleration-dependent velocity that makes their motion reduce relative to the cosmic framework. This additional acceleration due to cosmic pairings may be the same factor that causes the mass growth of compact bodies in the core of elliptical galaxies".
Another recently published article \cite{cc} examines the hypothesis that black holes can be a source of dark energy and guide the accelerated expansion of the universe. The authors assume in their paper that there are pairs of cosmic black holes with an index $k\approx 3$, which means that the mass of the black hole increases as $M_ {\rm BH}\propto a^k$ with the scale factor of the universe.
The authors tested this hypothesis with data from the James Webb Space Telescope (JWST), which detected more than 180 active galactic nuclei (AGN) and high-redshift quasars, and found that three of the AGNs have a mass ratio of $M_ {\star}/M_ {\rm BH}$ that does not match the prediction of the cosmic pair of black holes with $k = 3$ at the confidence level of $\sim 3\sigma$. They state that perhaps future JWST searches will lead to finding more AGNs in early-type host galaxies that have the above conditions and further confirm this discrepancy. Based on this concept, they conclude that the results of the above observations do not confirm the hypothesis that black holes are the source of dark energy.

\section{Original discussion}
One of the developing approaches to studying quantum gravity is the swampland program. Various contents of cosmology, such as inflation, physics of black holes, thermodynamics, dark energy, etc., have been studied through this approach to obtain information about the quantum nature of gravity. Black holes have always been a good playground for WGC, and recently extensive studies have been conducted. The WGC claims that extremal black holes must decay, so the result of this collapse includes elementary particles. But in the black hole version of the WGC, they may be black holes themselves. When black holes have electric and magnetic charges, the resulting daughter black hole will necessarily violate cosmic censorship, although this does not happen in all black holes. Recently, Alipor et al., \cite{5} claim that black holes that contain quintessence and cloud of strings reject the claim of cosmic censorship violation, and thus the cosmic censorship will be satisfied in such black holes. The WGC of the swampland program is considered as follows,
\begin{equation}\label{1}
\begin{split}
Q/M>1,
\end{split}
\end{equation}
where $Q$ and $M$ are the charge and mass of a black hole. Different versions of the WGC are considered in the literature, and one of the most important questions in the Swampland program arising from string theory is which of these versions can be considered the strongest and most significant for the WGC. As we stated earlier, one of the cases that we have for the WGC is the collapse of a charged black hole in an extremal state, which causes the production of particles with $q/m>1$ (where q and m are the charge and mass of the particles, respectively). In this case, we have,
\begin{equation}\label{2}
\begin{split}
(F_e=\frac{q^2}{r^2})>(F_g=\frac{m^2}{r^2}).
\end{split}
\end{equation}
The equation above shows that gravity is weaker than the repulsive force between particles resulting from the decay of an extremal black hole. It is important to note that we cannot apply the inequality $Q/M>1$ (where Q and M are the charge and mass of the black hole, respectively) to some black holes, such as charged black holes, as it would violate the weak cosmic censorship.
Recently, \cite{5} claimed that for a charged black hole in the presence of quintessence and a cloud of strings, the WGC is compatible with the weak cosmic censorship conjecture. Therefore, we rewrite the relation \eqref{2} for charged black holes as follows:
\begin{equation}\label{3}
\begin{split}
(F_e=\frac{Q^2}{r^2})>(F_g=\frac{M^2}{r^2}).
\end{split}
\end{equation}
According to equation \eqref{3} and the WGC (equation \eqref{1}), we find that such black holes can repel each other. Considering the explanations above and references \cite{5,6,7,8} as well as the concepts expressed in \cite{1,2,3,4}, where we briefly mentioned the results of their work, we would like to claim that certain black holes, such as those with quintessence, a cloud of strings, and perfect fluid, can be considered a factor for the expansion of the universe according to the mechanism of WGC.
Our main argument is that if two black holes surrounded by quintessence, a cloud of strings, or perfect fluid get close to each other, a repulsive force will be created between them with respect to the mechanism of the WGC. This force will cause them to move away from each other.
We can assume this repulsive relationship for a much larger number of the same family of black holes surrounded by quintessence or the cloud of string or perfect fluid get. As a result, the repulsive force that will exist between any two of these black holes leads to the moving away of these black holes to each other.
In fact, the movement of these black holes and their distance from each other, due to the repulsion force, may be an auxiliary factor in the further expansion of the universe. Of course, such an event does not occur in connection with ordinary black holes in relation to the WGC, but perhaps more evidence can be presented later. However, it seems at first glance, and considering that black holes are a good playground for the WGC, that more evidence for such an event can be obtained by checking other properties of black holes.\\\\
The metric of charged black hole in the presence of quintessence and cloud of string is as follows \cite{5},
\begin{equation}\label{eq4}
ds^2=f(r)dt^2-f^{-1}(r)dr^2-r^2(d\theta^2+\sin^2(\theta)d\varphi^2),
\end{equation}
with
\begin{equation}\label{eq5}
f(r)=1-b-\frac{2M}{r}+\frac{Q^2}{r^2}-\frac{ \alpha}{r^{3\omega_q+1}},
\end{equation}
where $Q$, $M$, $b$, $\alpha$, and $\omega_q$ are the electric charge, mass of the black hole, constant that accounts for the presence of the cloud of strings, positive normalization factor, and quintessential state parameter, respectively. In the following, we demonstrate using a diagram how the Weak Gravity Conjecture (WGC) and the Weak Cosmic Censorship Conjecture (WCCC) are compatible with our model.
\begin{figure}[h!]
 \begin{center}
 \subfigure[]{
 \includegraphics[height=5cm,width=5cm]{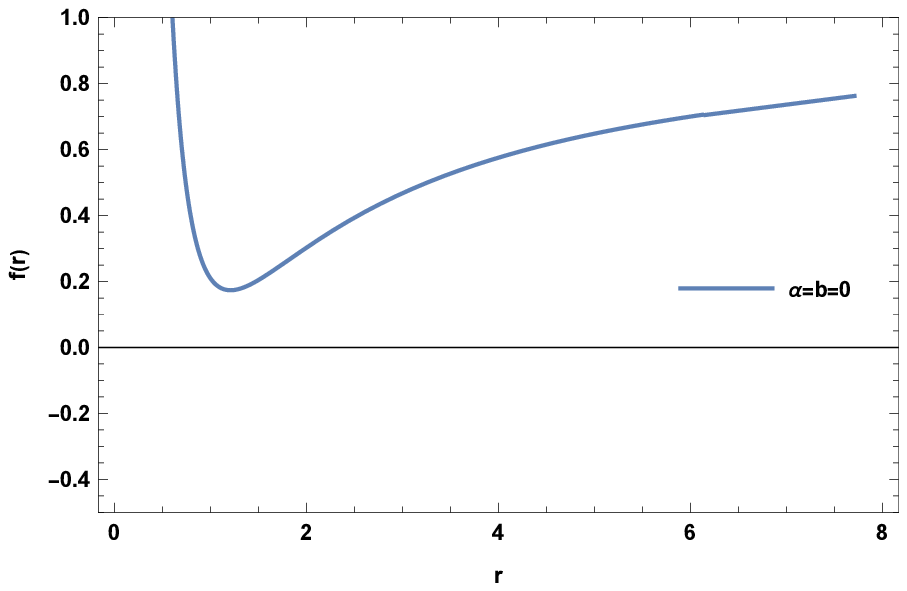}
 \label{1a}}
 \subfigure[]{
 \includegraphics[height=5cm,width=5cm]{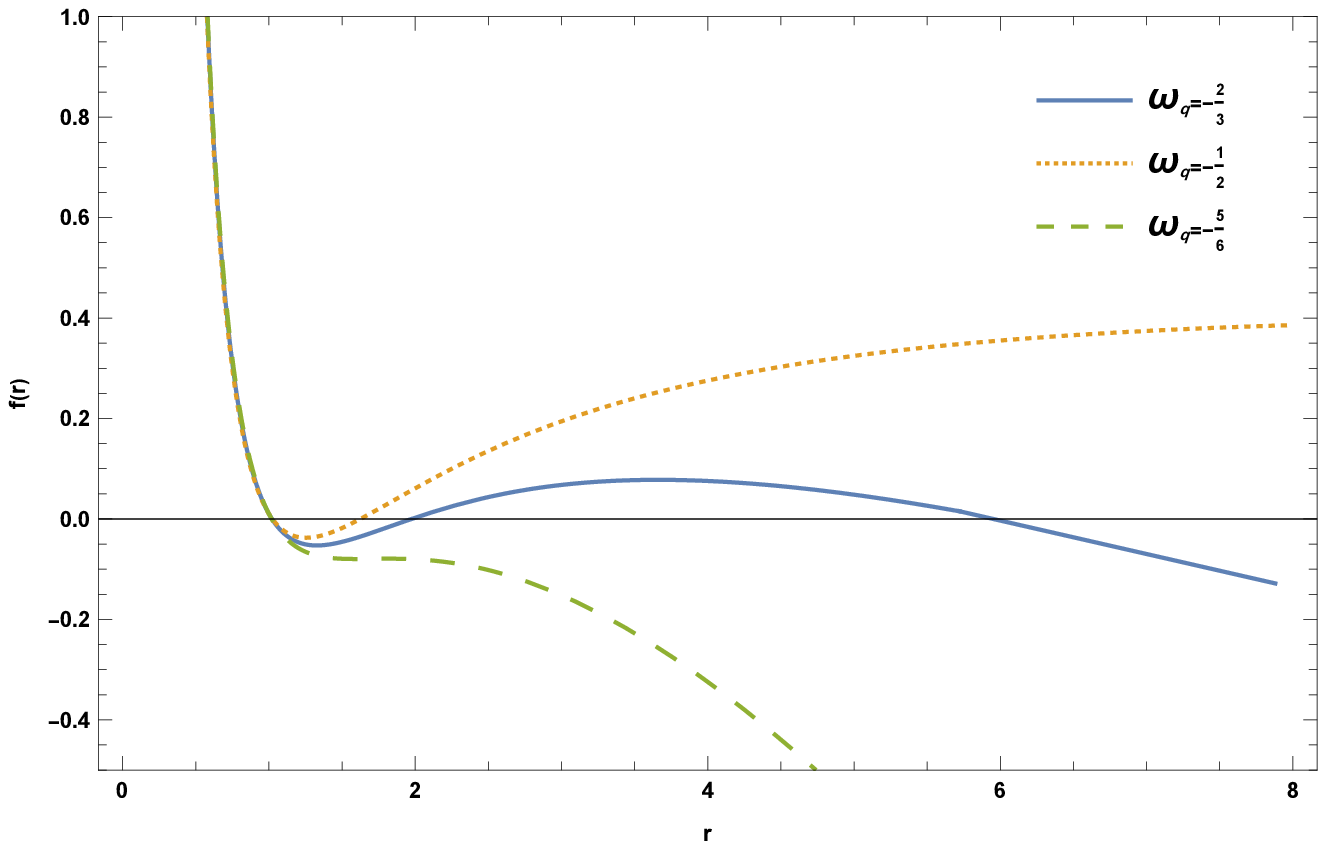}
 \label{1b}}
 \subfigure[]{
 \includegraphics[height=5cm,width=5cm]{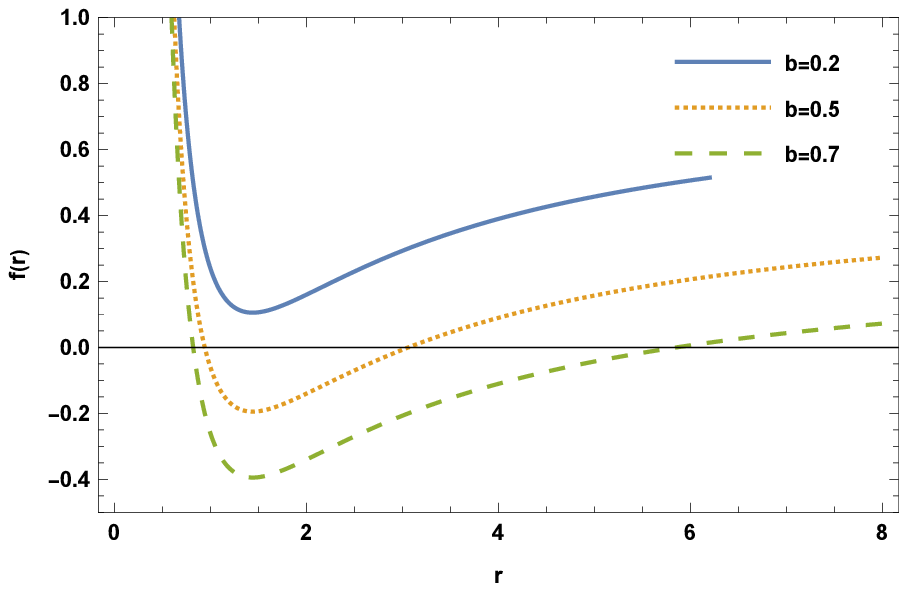}
 \label{1c}}
 \caption{\small{ (a) The metric function $f(r)$ for $M=1$, $Q = 1.1$, and $\alpha=b=0$ (Reissner–Nordström black hole). (b) The metric function $f(r)$ for different values of $\omega_q$. We choose
$M=1$, $Q = 1.1$, $\alpha=0.1$ and  $b=0.1$. (c) The metric function $f(r)$ for different values of $b$. We choose
$M=1$, $Q = 1.2$, and $\alpha=0$.}}
 \label{21}
 \end{center}
 \end{figure}
As we know, for a usual charged black hole, we will not have an event horizon when $Q/M>1$. Therefore, WGC is incompatible with WCCC for a usual charged black hole, as shown in Fig. 1a. However, for the charged black hole in the presence of quintessence and a cloud of strings (Fig. 1b) and the charged black hole in the presence of a cloud of strings (Fig. 1c), when $Q/M>1$, we will have event horizons for different values of $\omega_{q}$ and $b$. Therefore, the charged black hole in the presence of quintessence and a string cloud, despite satisfying WGC, will have an event horizon for the black hole, which leads to maintaining WCCC. In other words, for this type of black hole, WGC and WCCC are compatible with each other.
Furthermore, we will have for the $R-N$ black hole surrounded by perfect fluid dark matter\cite{9},
 \begin{equation}\label{6}
\begin{split}
g(r)=1-\frac{2M}{r}+\frac{Q^2}{r^2}-\frac{\alpha}{r}\ln{\frac{r}{\alpha}},
\end{split}
\end{equation}
where $M$ and $Q$ are the mass and charge of the black hole, respectively, and $\alpha$ is the parameter for perfect fluid dark matter.
 \begin{figure}[h!]
 \begin{center}
 \includegraphics[height=5cm,width=8cm]{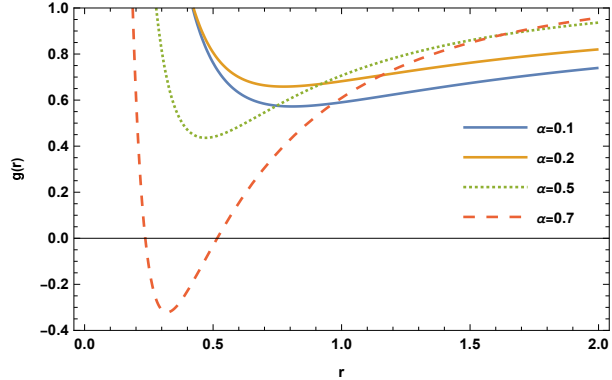}
 \label{2}
 \caption{\small{ The metric function $g(r)$ for $M=0.5$, $Q =0.6$, and  different values of $\alpha$.}}
 \label{21}
 \end{center}
 \end{figure}
As shown in Fig. 2, there will be an event horizon for the black hole in the $Q/M>1$ state for specific values of ($\alpha$).
\textit{Therefore, for these black holes, the WGC and the WCCC can be compatible with each other. As explained in detail above, by considering charged black holes surrounded by quintessence, string cloud, and perfect fluid, it is possible to introduce a repulsion force for similar family black holes by satisfying the condition of the WGC. Therefore, for the aforementioned black holes, there exist certain values of free parameters that make the WGC compatible with the WCCC, and we also observe a repulsion force. However, this relationship cannot be fully established for the $R-N$ black hole because the WCCC is violated in such black holes, as shown in Figure 1a. Therefore, by considering different types of black holes and maintaining these conditions, it may be possible to discuss and categorize black holes from the same family as an auxiliary factor, albeit a weak one, in the expansion of the universe.}
\section{Conclusion}
Our work in this article is presented in two different formats, which are expressed and used in the same way. They can be checked as follows: \\
From one point of view, we present a new hypothesis and examine the evidence and conditions raised in this case. In the second form, we simultaneously present evidence of the application, impact, and motivation of a new hypothesis in quantum gravity as a tool for our work. We demonstrate how accurate this hypothesis can be and show that the results of its application are in line with the usual and known methods.\\\\

\textit{The first format}:\\
When it comes to the acceleration and expansion of the universe, the first and most important hypothesis that draws attention is an unknown and mysterious concept called dark energy. It is a quantity with an unknown origin that abundantly shows its influence and presence in our results and relationships, especially in terms of the accelerating universe.
The debate regarding the origin and influence of this mysterious parameter is still ongoing, and there is no certainty in this regard. However, as the human mindset always attempts to unify scientific methods for studying phenomena, in the case of the accelerated universe, the prevailing idea among most physicists is that dark energy is the sole effective factor in producing this acceleration.
In this article, we propose another suggestion, which is that perhaps dark energy is not the only factor contributing to the acceleration of the universe. One possible factor could be the formation of multiple pairs of repulsive black holes that are stably scattered throughout the universe, leading to an increase or maintenance of acceleration through mutual repulsion.
It is possible to go even further and claim that perhaps dark energy is not actually unified. This apparent unity may be due to the type of glasses we have used. If we change our perspective a little, it may become clear that dark energy is actually made up of a large number of factors, just like white light that needs a prism to reveal its underlying colors.\\\\

\textit{The second format}:\\
The nascent hypothesis of the WGC has not yet received the general favor of the physics community as it should. Perhaps the most important reason for this is the lack of verification by concrete physical phenomena and experiments.
The challenging conditions for practical testing of this hypothesis, which require very high temperatures within very small ranges, have caused physicists to use it cautiously. However, new results and experiments every day provide more evidence of the validity of this hypothesis in examining cosmic structures.
In this article, we introduced and used the results of this hypothesis in our work to show how accurate and usable this hypothesis can be.
Matching the justification or conjectures of this hypothesis with modern experimental results obtained, especially for the study of cosmic structures which is clearly evident in our work, can be a suitable step to pay more attention to this strong and practical hypothesis.\\\\
The *Weak Gravity Conjecture* (WGC) is a hypothesis about quantum gravity that states that gravity is always the weakest force in any consistent theory of quantum gravity. This means that there must exist some particles that have a higher charge-to-mass ratio than an extremal black hole and therefore can escape its gravitational attraction.
The *repulsive force conjecture* (RFC) is a hypothesis that states there must exist particles that are self-repulsive, meaning that two copies of the same particle repel each other when they are far apart. This implies that the gauge force between them is stronger than the gravitational force.
The connection between the WGC, black holes, and the RFC is that both conjectures are motivated by consistent arguments involving black holes. For example, the WGC can be seen as a way to avoid having stable sub-extremal black holes, which would violate charge conservation and thermodynamics. The RFC can be seen as a way to avoid having super-extremal black holes, which would have negative mass and repel other objects.
However, the precise formulations and implications of these conjectures are not fully understood, especially in the presence of massless scalar fields (moduli) that can affect the long-range forces. There are also different versions of these conjectures, such as strong forms that require an infinite tower of particles with increasing charge-to-mass ratios or repulsive forces. Therefore, more research is needed to test and refine these conjectures in various examples of quantum gravity, such as string theory.\\\\
There are possible connections between the WGC and the expansion of the universe. One connection is that the WGC may imply a bound on the cosmological constant, which is the parameter that drives the accelerated expansion of the universe. If the cosmological constant is too large, it may violate the WGC by making gravity stronger than any other force. Therefore, the WGC may constrain the possible values of the cosmological constant and hence the rate of expansion of the universe.
Another possible connection is that the WGC may imply a relationship between inflation and quantum gravity. Inflation is a hypothetical period of rapid expansion in the early universe that can explain some of its features, such as its flatness and homogeneity. However, inflation also requires a scalar field with a very flat potential, which may be difficult to realize in a consistent theory of quantum gravity. The WGC may provide a criterion for selecting viable models of inflation that are compatible with quantum gravity.\\\\
Recently, it was shown in \cite{1} that black holes are a source of dark energy. In \cite{2,3,4,cc}, the truth or falsity of this concept has been discussed.
We have shown that a charged black hole, in the presence of quintessence, a string cloud, and a perfect fluid, can produce a repulsive force with respect to the WGC mechanism that overcomes the gravitational force.
\textit{We explained that these repulsive forces, present in a population of black holes belonging to the same family, can be considered as an auxiliary factor for the expansion of the universe}.
Therefore, it is possible to examine other black holes from this point of view and look for more evidence of this idea. Of course, we hope that in the future, through more observations, stronger evidence for the connection between the above concepts will be obtained.

\end{document}